\input epsf.sty
\documentstyle[prd,preprint,aps]{revtex}
\newcommand{\Caption}[1]{\caption{{\it #1\\~\\}}}
\newcommand{\incfig}[1]{\centerline{\epsfysize=3.5in \epsfbox{#1}}}
\newcommand{\incfif}[1]{\centerline{\epsfysize=3.0in \epsfbox{#1}}}

\tighten
\begin{document}

\title{New Orthogonal Body--Fitting Coordinates
       for Colliding Black Hole Spacetimes}
\author{Peter Anninos${}^{(1)}$, Steven Brandt${}^{(2)}$, Paul
Walker${}^{(2,3)}$}
\address{
${}^{(1)}$ National Center for Supercomputing Applications,
Beckman Institute, 405 N. Mathews Avenue, Urbana, Illinois, 61801 \\
${}^{(2)}$ Max--Planck--Institut f\"ur Gravitiationsphysik
(Albert--Einstein--Institut), Schlaatzweg 1, 14473 Potsdam, Germany \\
${}^{(3)}$ Department of Physics,
University of Illinois, Urbana, Ilinois 61801
}

\date{\today}
\maketitle

\begin{abstract}
We describe a grid generation procedure designed to 
construct new classes of orthogonal coordinate systems for
binary black hole spacetimes. The computed coordinates offer an
alternative approach to current methods, in addition to
providing a framework for potentially more stable and accurate
evolutions of colliding black holes. 
As a particular example, we
apply our procedure to generate appropriate numerical grids to
evolve Misner's axisymmetric initial data set representing
two equal mass black holes colliding head--on.
These new results are compared with
previously published calculations, and we find
generally good agreement
in both the waveform profiles and total radiated energies
over the allowable range of separation parameters.
Furthermore, because no specialized treatment of the coordinate
singularities is required, these new grids are more
easily extendible to unequal mass and spinning black hole
collisions.
\end{abstract}
\pacs{PACS numbers: 04.25.Dm, 04.70.-s, 95.30.Sf }

\section{Introduction}
\label{sec:introduction}
After the first attempt in 1964 
by Hahn and Lindquist~\cite{Hahn} to solve the problem of colliding
black holes, a long--term and relatively 
successful program was initiated in the late 1960s
by DeWitt et. al. (see DeWitt, \v{C}ade\v{z}, Smarr, 
and Eppley \cite{Cadez,Smarr1,Eppley,DeWitt},
which we henceforth collectively refer to as DCSE).
Their efforts were concentrated on simulating
the axisymmetric head--on collision of two equal mass black holes using
the time symmetric Misner data~\cite{Misner} for initial
conditions. This data set possesses the ``double Einstein--Rosen
bridge''~\cite{Einstein} topology in which two asymptotically flat
sheets are joined by two throats representing non-rotating 
black holes. The throats are two--spheres that are invariant under
an isometry operation identifying both sheets and, from a
numerical standpoint, serve as a boundary across which a
particular form of Neumann boundary condition must be imposed on 
the evolved quantities to preserve the isometry.

DCSE developed the 2D \v{C}ade\v{z} \cite{Cadez} coordinate system
to numerically integrate and evolve the
axisymmetric Einstein equations.
The \v{C}ade\v{z} coordinates are curvilinear
body--fitting coordinates that 
conform to the two black hole throats and become
spherical at asymptotic infinity. The
advantages of this coordinate system (and any other appropriate
body--fitting system) include: (1) simplified boundary conditions at the
black hole throats, (2) natural spherical--like grids which mimic the
background Schwarzschild geometry that the
solutions approach at late times, (3) simplification of waveform
extraction calculations in the wave zone, and (4) exponential stretching
of the grid in the radial direction which covers enough of
the spacetime that asymptotic flatness can be applied to the conformal
metric components at the outer grid edges.
However, these conveniences are offset somewhat by the singular
saddle point introduced by the coordinate transformation at the
origin midway between the two black holes. DCSE attempted
to deal with the
saddle point by evolving the cylindrical (not \v{C}ade\v{z}) metric components
everywhere on the \v{C}ade\v{z} grid and performing discretizations using
chain rule derivatives. Also, because black hole spacetimes can generate
pathologically steep delta--function--like peaks in the metric components
(assuming singularity avoiding time slices), numerical
evolutions can quickly become unstable even if
the inherent symmetries are exploited in choosing
the appropriate numerical grid and the evolved variables.
Additionally, it is well
known~\cite{Smarr1,Eppley} that an axis instability can be triggered in
numerical evolutions of axisymmetric spacetimes. 
For all of these reasons, the
computations of DCSE remained uncertain with error bars of
order 100\%~\cite{Smarr2}. 

Recently, Anninos, Hobill, Seidel, Smarr and
Suen~\cite{Anninos1,Anninos2,Anninos3} (henceforth referred
to as Papers I, II and III respectively) improved
upon the DCSE calculations by introducing a shift vector to
diagonalize the \v{C}ade\v{z} metric and to evolve these components on the
\v{C}ade\v{z} grid, thus suppressing the axis instability which is
especially sensitive to the off--diagonal elements.
The problem with the coordinate singularity 
at the origin was treated by
constructing a cylindrical coordinate ``patch'' to cover regions near
the saddle point. The results of Anninos et al.
are accurate to a few percent in the dominant wave signal for black
holes which are initially 
placed at close to moderate separation distances.
For larger initial separations (greater than about $10 M$,
where $M$ is the single black hole mass), 
the evolutions become increasingly
more inaccurate due in part to the fact that the saddle point remains
within the causally connected regions of spacetime for longer
periods of time as the black
holes take longer to collide at the origin. If the black
holes are initially separated by even greater distances
(more than about $20 M$), the evolutions
become unstable and break down altogether.

In short, the goal of long--term stable 
and accurate computations of highly separated colliding black
holes has yet to be achieved.
Here we propose two new classes of
orthogonal body--fitting coordinate systems that are well--suited to the
axisymmetric 
geometry of two colliding black holes, and that appear promising to improve
upon current calculations. The new coordinates remove the singular
saddle point from its obtrusive position at the origin and thus offer a
potentially cleaner and more stable method of solution.  
Because we require spherical
coordinate lines at the throats and in the far field, singular points
are unavoidable. However, the saddle point singularities 
can be relocated along the $z$--axis to either
the south or north pole of the top hole (and thus
the north or south pole of the bottom hole, respectively),
assuming the two holes are aligned vertically along the $z$--axis.
We refer to the former system where the singularities
face the opposite black hole as the class I system:
class II refers to the latter case with the singularities
facing away from the opposing throats.

We are thus able to construct
two new classes of coordinate systems, replacing the
single singularity at the origin with a pair of 
saddle singularities on the
throats. An advantage of transplanting the coordinate singularities
to the throats prevents the black holes from
``crashing'' into a saddle point during the numerical evolutions.
Instead, we implement a singularity avoiding lapse function 
that is zero on the throat,
so that the evolution along the throat freezes in time.  In
addition, the (maximal) time slicing used in
the evolutions will rapidly absorb coordinate
lines into the event horizon and the saddle points will fall further
inside the causally disconnected portion of the evolved spacetime
where the lapse collapses exponentially to zero, providing
an additional stabilizing element.

The following sections describe our hybrid analytic/numerical
procedure to generate discrete grids for axisymmetric binary
black hole systems.
Sec.~\ref{sec:radial} introduces grid generating techniques 
in general and
describes the particular analytic prescriptions that we developed for 
specifying one of the coordinates. Sec.~\ref{sec:angular}
describes the numerical calculation of the second coordinate orthogonal
to the first, as well as
the techniques used to compute the Jacobian matrix required for the
coordinate transformations, and the
tests which can be used to monitor the accuracy in which
the coordinate systems and Jacobians are constructed.
Results from actual numerical
evolutions are presented in Sec.~\ref{sec:applications} and compared
with previously published calculations. 
We summarize our results in Sec.~\ref{sec:conclusion}.

\section{Generating Body--fitting coordinates: The Specified Coordinate}
\label{sec:radial}

A general and common method of 
constructing body--fitting coordinates is to let the
curvilinear coordinates ($\eta$, $\xi$) satisfy the following elliptic
partial differential equations~\cite{Thompson}
\begin{mathletters}
\label{eqn:laplace}
\begin{eqnarray}
\partial_{xx}\xi  + \partial_{yy}\xi  &=& P(\eta,~\xi)~, \\
\partial_{xx}\eta + \partial_{yy}\eta &=& Q(\eta,~\xi)~,
\end{eqnarray}
\end{mathletters}where 
$P(\xi,\eta)$ and $Q(\xi,\eta)$ are generating functions that can
be used to adjust the behavior of the curvilinear coordinates
in the physical domain. 
Because the boundaries are typically irregular in Cartesian
coordinates (or cylindrical coordinates in our black hole work),
and the ($\eta$, $\xi$)
coordinates are uniform in the transformed plane, it is
desirable to carry out the computations in the transformed plane,
switching the dependent and independent variables. Equations
(\ref{eqn:laplace}) can be inverted to yield
\begin{mathletters}
\label{eqn:thomp}
\begin{eqnarray}
\alpha  \partial_{\xi\xi}x  -2\beta \partial_{\eta\xi}x +
\gamma  \partial_{\eta\eta}x  
+ J^2\left(P \partial_{\xi}x + Q \partial_{\eta}x\right) &=& 0~, \\
\alpha \partial_{\xi\xi}y  -2\beta \partial_{\eta\xi}y +
\gamma \partial_{\eta\eta}y  +J^2\left(P \partial_{\xi}y 
+ Q \partial_{\eta}y \right) &=& 0~,
\end{eqnarray}
\end{mathletters}where
$\alpha=\partial_\eta x^2 + \partial_\eta y^2$, 
$\beta = \partial_\xi x \partial_\eta x + \partial_\eta y \partial_\xi y$,
$\gamma=\partial_\xi x^2 + \partial_\xi y^2$, and 
$J = \partial_\xi x \partial_\eta y - \partial_\eta x \partial_\xi y$.
The problem of grid generation then reduces to solving
a coupled set of nonlinear elliptic partial differential equations
subject to the appropriate boundary conditions, which can
themselves be intrinsically coupled in a complicated manner.
Although the elliptic solution approach is a general 
and straight--forward one,
other methods such as conformal mapping, algebraic
transformations, and hyperbolic solutions of
partial differential equations have also
been developed (see \cite{Thompson} for a review
of grid generation techniques). 

In this paper, we present an altogether different and much 
simplified procedure that does not require one to solve 
coupled nonlinear elliptic equations.
The idea rests on the notion that the grid generation process
is greatly simplified if one of the coordinates can be
specified analytically. Taking this coordinate
as the one aligned asymptotically with 
either the radial or the angular
direction, we show in this section how to construct a
natural ``specified'' coordinate for three different classes of grids
as characterized by the location of the singularity:
class I (II) with two singular saddle points, one on each
of the throats on the axis closest to (furthest from) the origin;
and class III which are \v{C}ade\v{z}--like
coordinates with a single singular point at the origin $\rho=z=0$.
We note that the function one chooses for the specified
coordinate must solve the necessary boundary conditions, but is
not restricted to satisfy any particular elliptic equation.

Before we continue,
a few notational comments are in order. Four
axisymmetric coordinate systems are utilized in this paper: 
the combination $(\rho,z)$ refers to standard cylindrical coordinates, 
($r$, $\theta$) denotes standard polar coordinates,
($\zeta$, $\xi$) are the 
radial-- and angular--like body fitting coordinates, and 
($\eta$, $\xi$) represent the logarithmic radial-- and angular--like
body fitting coordinates (note the two $\xi$ coordinates are
identical). Finally we characterize the location of the two black
holes by their radius $a_1$ ($a_2$) and
vertical distance from the origin along the $z$--axis $z_1$ ($z_2$),
where $z_1$ ($z_2$) represents the {\it absolute} distance of the
throat center from the origin, and the
upper (lower) black hole is denoted by the subscript 1 (2).

\subsection{The Class I System}
\label{sec:classI}

The class I coordinate system has a coordinate 
(saddle--type) singularity where
each of the throats meet the axis closest to the origin 
(see Fig. \ref{fig:clI_misgrid}).
A singularity is generated at these points by requiring that a
radial--like coordinate $\zeta$ be zero on both of the throats and the
entire section of the axis connecting the throats. Lines of constant
$\zeta$ will then transform from ``peanut''--like 
surfaces near the throats to radial circles at infinity.
The two throats and the axis between the throats
in this case make up the line $\eta = 0$. The 
axis above (below) the top (bottom) hole is the angular
coordinate value $\xi = 0$ ($\pi$).
In the equal mass case, the equator is the line $\xi=\theta =\pi/2$.
The boundary conditions are shown in Fig.~\ref{fig:clI_bound}  
and Fig.~\ref{fig:clI_schematic} in the
($\rho, z$) and ($\zeta, \xi$) planes respectively.

\firstfigfalse
\vbox{
\figure{
\incfig{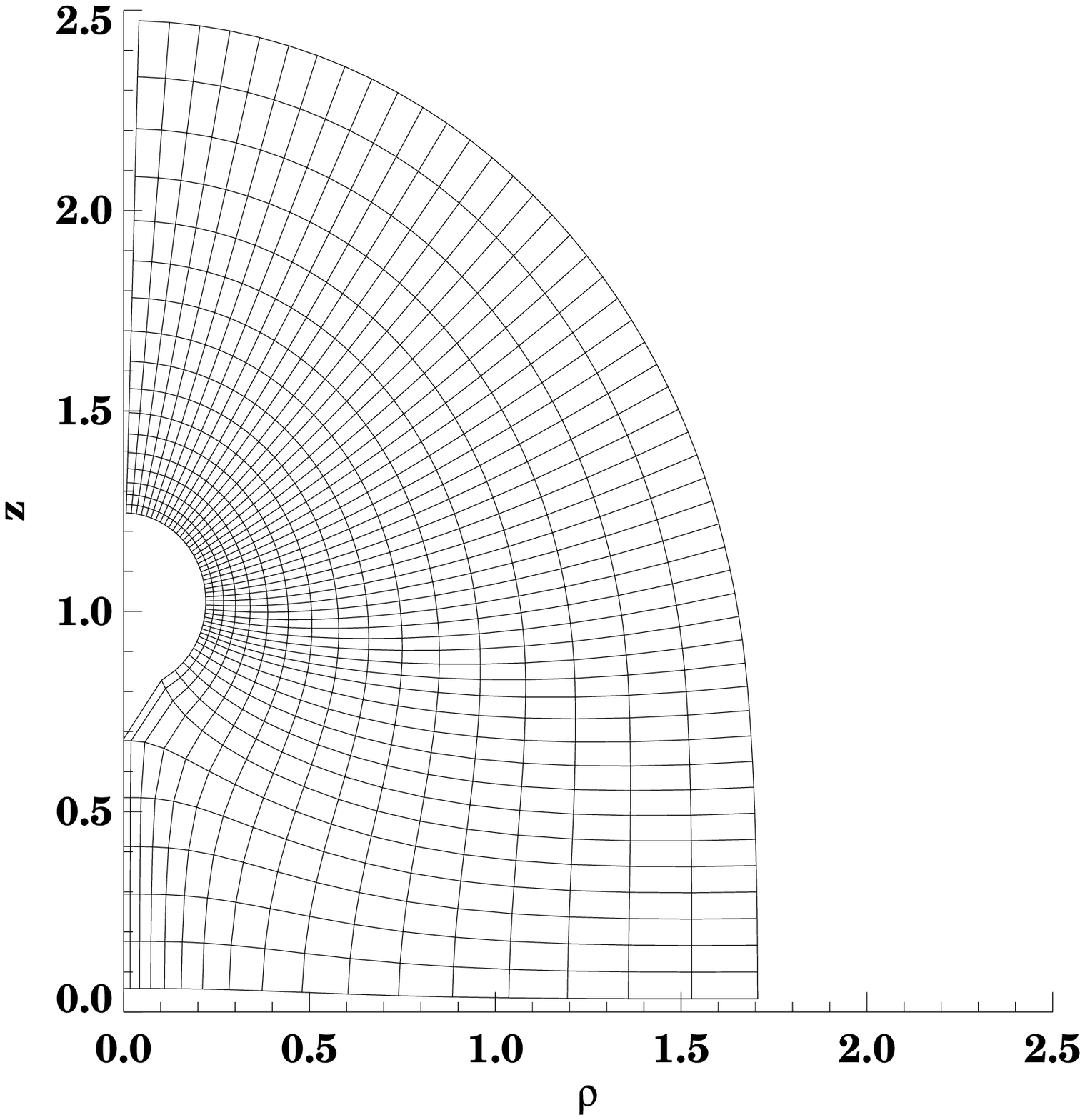}
\Caption{
Class I coordinates for the $\mu = 2.2$ Misner data.}
\label{fig:clI_misgrid}
}
}

An exponentially stretched radial coordinate satisfying the
appropriate boundary conditions can be generated by first
defining two radial distances from the two throats
\begin{eqnarray}
d_1 &=& \sqrt{\rho^2 + (z-z_1)^2} - a_1 , \\
d_2 &=& \sqrt{\rho^2 + (z+z_2)^2} - a_2 ,
\end{eqnarray}
and a third distance measure 
\begin{eqnarray}
d_3 &=& \frac{1}{2}\left[\sqrt{\rho^2 + (z-z_1+a_1)^2} +\right.\nonumber\\
  &&\left.\sqrt{\rho^2+(z+z_2-a_2)^2}-(z_1+z_2-a_1-a_2)\right] ,
\end{eqnarray}
defining an ``elliptic'' radial coordinate that is zero on the
segment along the $z$--axis extending from $z=z_1$ to $z=-z_2$.
(We note that the square root radicals implicitly refer to the
absolute or positive root.)
The actual curvilinear ``radial'' coordinates, $\zeta$ and $\eta$,
are then constructed from these distance components by
\begin{equation}
\zeta = \frac{(2 \kappa_1 + 1)d_1 d_2 d_3 }
             {d_1 d_2 + \kappa_1(d_1 d_3 + d_2 d_3)} ,
\label{eqn:zeta_I}
\end{equation}
and 
\begin{equation}
\eta = \sinh^{-1} \left(\frac{\zeta}{\kappa_2}\right).
\label{eqn:eta_I}
\end{equation}

\vbox{
\figure{
\incfig{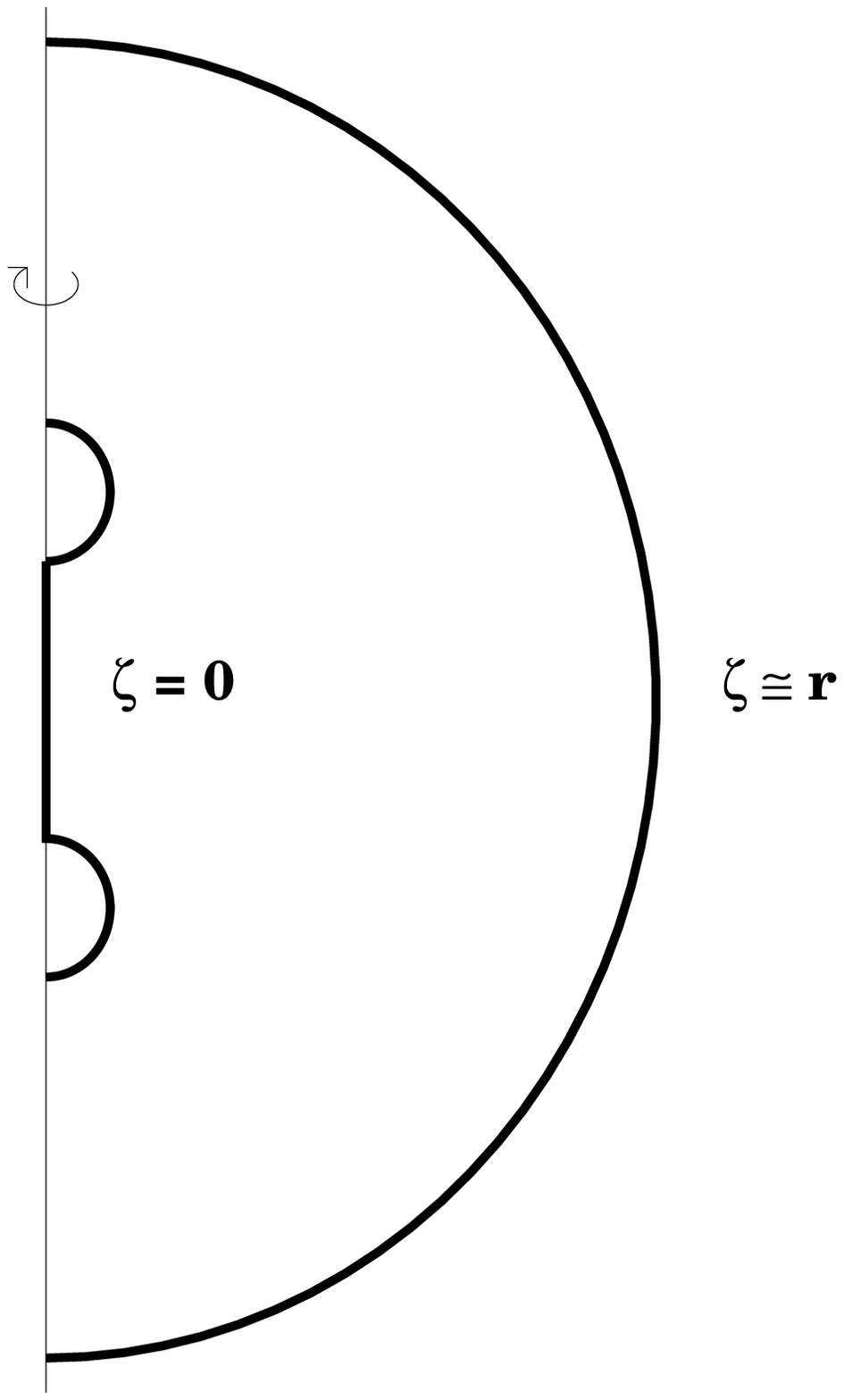}
\Caption{
The ``radial'' ($\zeta$) boundary conditions for the class I
coordinate system.}
\label{fig:clI_bound}
}
}
\vbox{
\figure{
\incfig{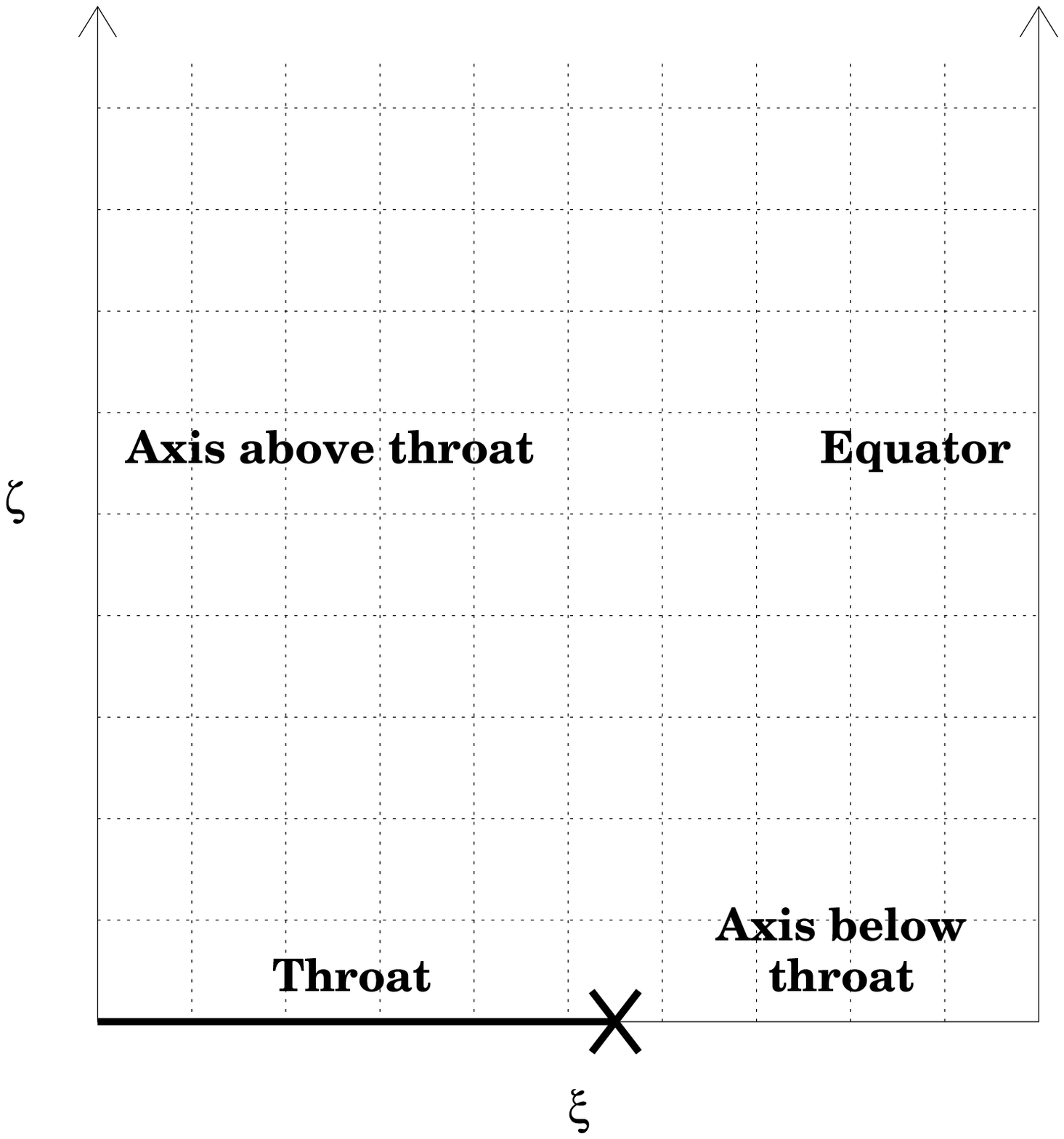}
\Caption{
The class I schematic mapping of the rectangular
$(\zeta,\xi)$ domain in relation to the throat, axis, and equator.
The cross indicates the location of the coordinate singularity at the
intersection of the throat and axis closest to the equator.}
\label{fig:clI_schematic}
}
}

\noindent
Both are zero on the ``spectacle'' boundary of
Fig.~\ref{fig:clI_bound} by construction.
Equations (\ref{eqn:zeta_I}) and (\ref{eqn:eta_I}) also have the proper
behavior in the asymptotic limit where both $\rho$ and $z$
tend to infinity, ie., $\zeta \rightarrow r$ and
$\eta \rightarrow \log(r/\kappa_2)$.
The parameters $\kappa_1$ and $\kappa_2$
are introduced so the coordinate systems can be ``tuned''
to improve the stability of evolutions. In particular, $\kappa_1$
controls the relative importance between the ``elliptic'' 
radius ($d_3$) and the two distance--from--throat radii ($d_1$ and $d_2$),
and affects the shape of the
grid close to the axis between the holes. By adjusting $\kappa_1$,
the grid can be pushed away from or drawn closer to the axis in this
region, effectively increasing or decreasing the resolution 
between the holes in the $\rho$ direction.
$\kappa_2$ controls the asymptotic behavior of the
coordinates in the far zone. Typical values for evolutions depend on
the geometry of the system, and range from
($\kappa_1$, $\kappa_2$) = (1, 2) for the $\mu=1.2$ case,
to (0.7, 2) for the $\mu=2.7$ case, where $\mu$ is the
Misner parameter defined in Sec.~\ref{subsec:black}.

\subsection{The Class II System}
\label{sec:classII}

The class II coordinate system has a  saddle point where each
of the throats meet the axis furthest from the origin 
(see Fig. \ref{fig:clII_misgrid}).
A singularity is generated at these points by requiring that
an angular--like coordinate $\xi$ be zero ($\pi$) along the throat 
of the top (bottom) black hole and along the axis above (below)
the top (bottom) throat. The coordinate $\xi$ also asymptotes to
the polar angular coordinate $\theta$.
The north (south) throat in this case is a line of constant $\xi = 0$
($\pi$) between $\zeta=0$ and some finite $\zeta$ value. 
The axis above (below)
the north (south) throat is on the same constant $\xi$ line as the
corresponding throats. The axis between the two throats is the line
$\zeta = 0$. In the equal mass case,
the equator is the line $\xi = \pi/2$. 
The boundary conditions and mapping are shown in
Fig.~\ref{fig:clII_bound} and Fig.~\ref{fig:clII_schematic}
in the ($\rho$, $z$) and ($\zeta$, $\xi$) planes respectively.

\vbox{
\figure{
\incfig{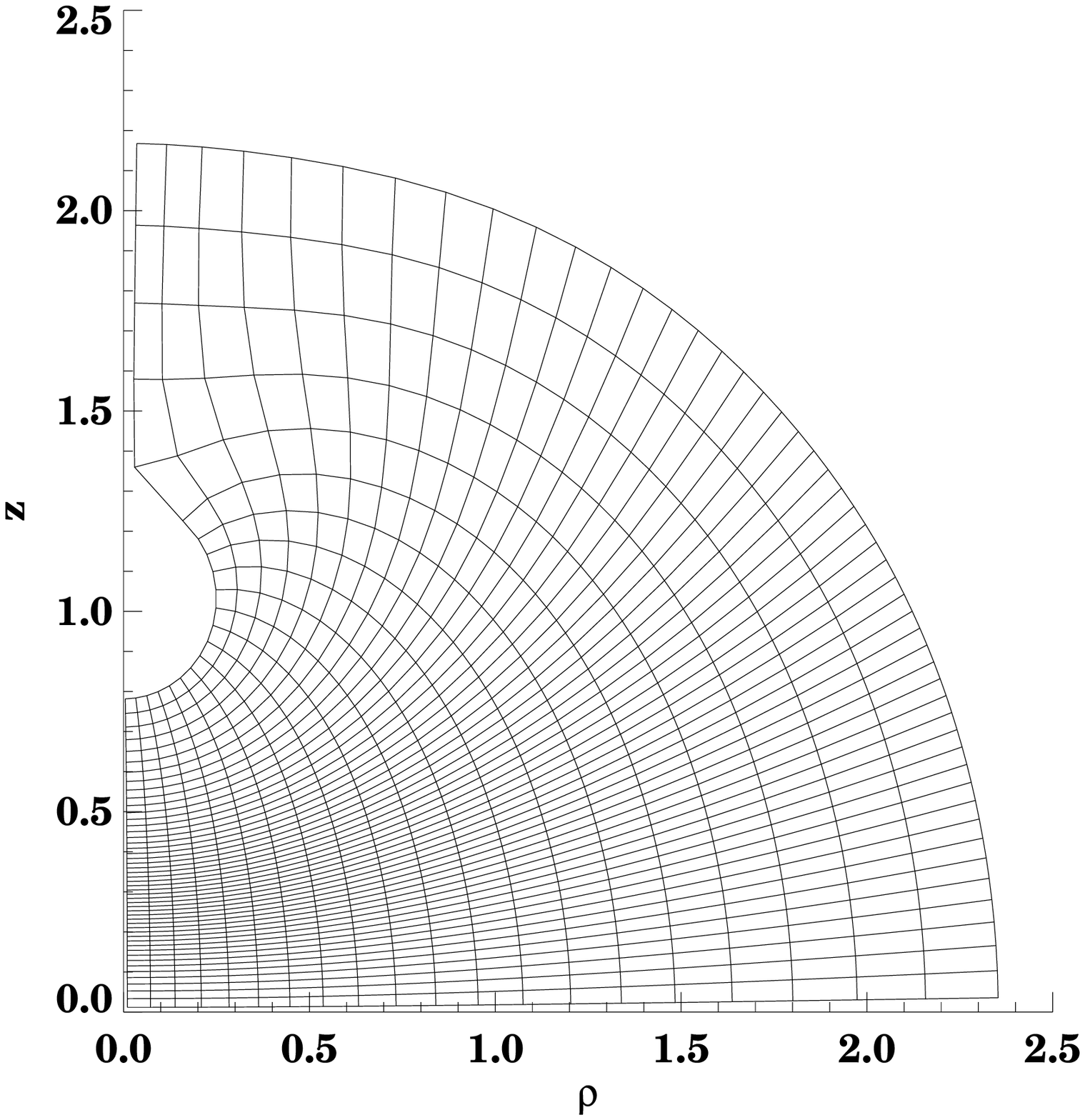}
\Caption{
Class II coordinates for the $\mu = 2.2$ Misner data.}
\label{fig:clII_misgrid}
}
}

\vbox{
\figure{
\incfig{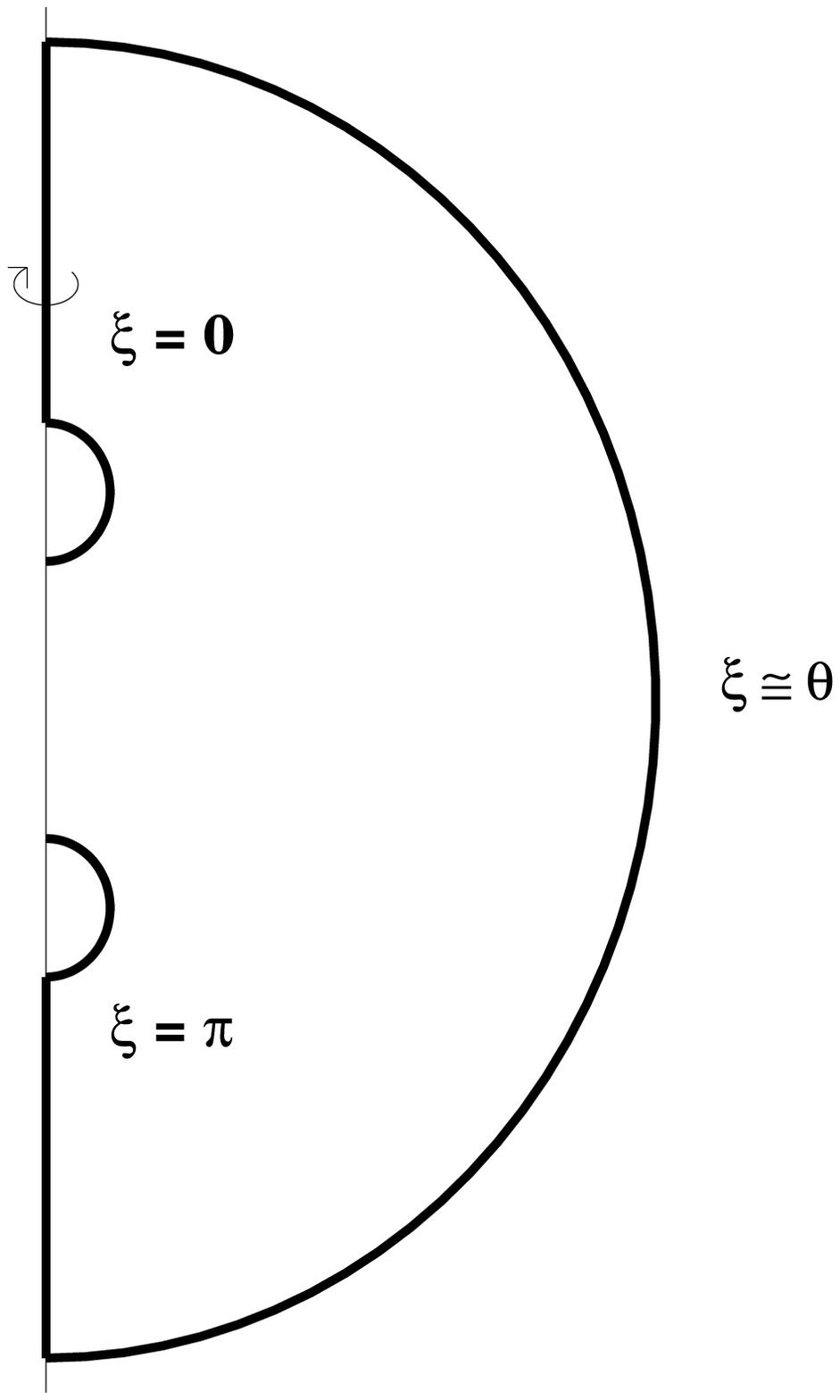}
\Caption{
The ``angular'' ($\xi$) boundary conditions for the class II
coordinates.}
\label{fig:clII_bound}
}
}
\vbox{
\figure{
\incfig{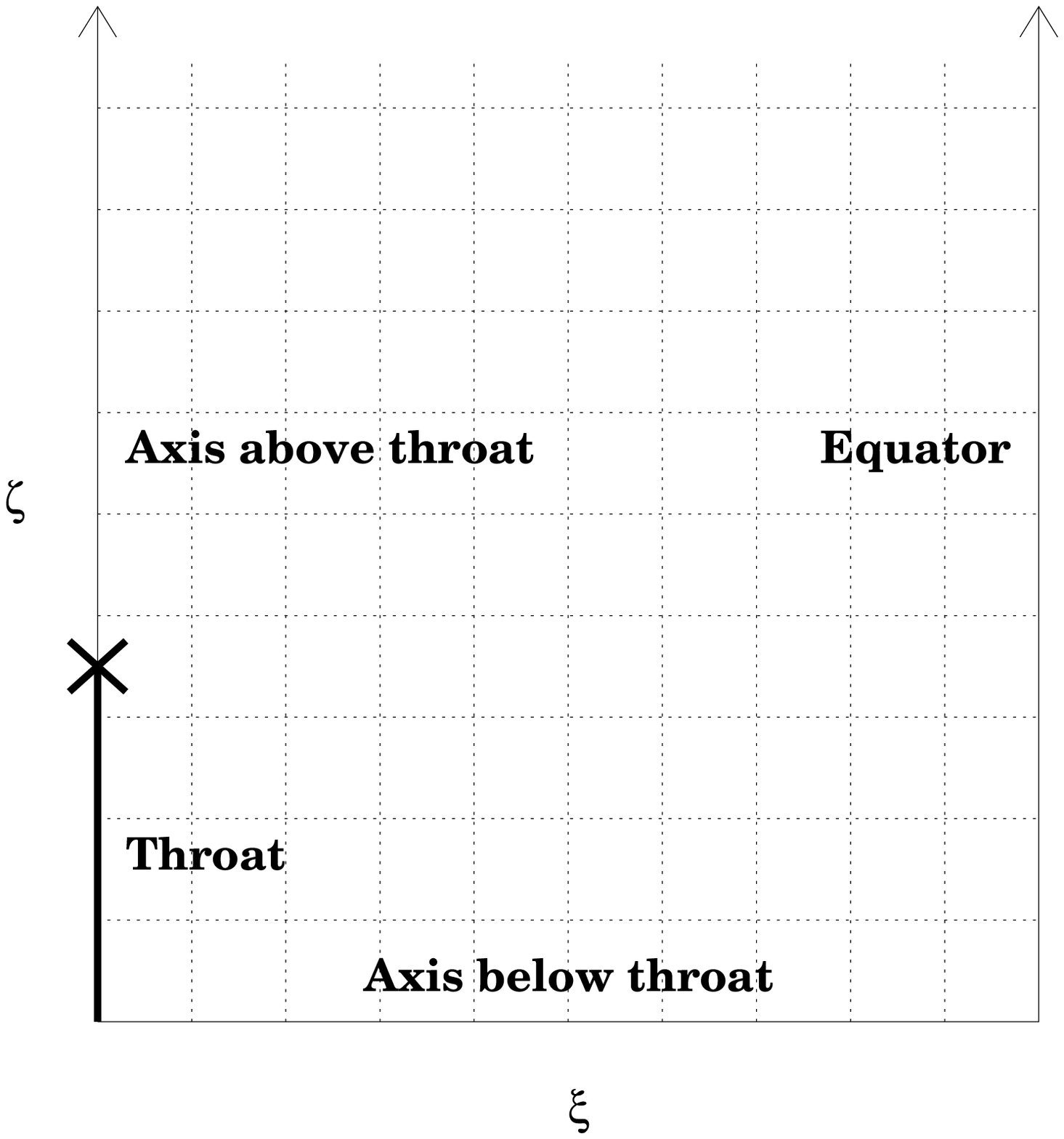}
\Caption{
The class II schematic mapping of the rectangular
$(\zeta,\xi)$ domain in relation to the throat, axis, and equator. The
cross indicates the location of the coordinate singularity
at the intersection of the throat and axis furthest from the equator.}
\label{fig:clII_schematic}
}
}

The class II system can be generated by first defining two
radial distances similar to the class I case
\begin{eqnarray}
d_{1a} &=& \rho^2 + (z-z_1)^2 - a_1^2 , \\
d_{2a} &=& \rho^2 + (z+z_2)^2 - a_2^2 ,
\end{eqnarray}
and two ``hyperbolic'' radial coordinates
\begin{eqnarray}
d_{1b} &=& -\sqrt{\rho^2+(z+z_2+a_2)^2} \nonumber\\
        &+&\sqrt{\rho^2+(z-z_1-a_1)^2}+(z_1+z_2+a_1+a_2) ,\\
d_{2b} &=& +\sqrt{\rho^2+(z+z_2+a_2)^2} \nonumber\\
        &-&\sqrt{\rho^2+(z-z_1-a_1)^2}+(z_1+z_2+a_1+a_2) ,
\end{eqnarray}
which go to zero along the axis above and below the holes respectively.
These coordinates can be combined to construct a function that
is zero on the top hole's throat and along the axis above it ($d_1$),
and a function that is zero on the bottom hole's throat and along
the axis below it ($d_2$)
\begin{eqnarray}
d_1 &=& \frac{d_{1a} d_{1b}}{\rho^2+(z-z_1)^2} , \\
d_2 &=& \frac{d_{2a} d_{2b}}{\rho^2+(z+z_2)^2} .
\end{eqnarray}
The two functions $d_1$ and $d_2$ are in turn combined to create
a common function $f$ that approaches $+\infty$ as $d_1 \rightarrow 0$,
and $-\infty$ as $d_2 \rightarrow 0$
\begin{equation}
f = (a_1+a_2+z_1+z_2)\left(\frac{1}{d_1}-\frac{1}{d_2}\right) .
\end{equation}
Finally, the function $f$ is converted to an angular--like coordinate by
\begin{equation}
\xi = {\rm ArcCot}\left(\frac{{\rm sign}(f)}{\sqrt{2}} 
      \sqrt{-1+\sqrt{1+f^2}}\right).
\label{eqn:class2_xi}
\end{equation}
The angular coordinate in Eqn. (\ref{eqn:class2_xi}) has the desired
properties of being zero on the throat of the top hole
and along the axis above it, $\pi$ on the bottom hole and along the
axis below it, and approaches the normal spherical
coordinate $\theta$ in the asymptotic limit $r \rightarrow \infty$.

In the class I system, the value of the second coordinate $\xi$
is specified on the outer grid edge to be the angular polar 
coordinate $\theta$. There is slightly more freedom in class II
for specifying the second coordinate $\eta$ at small $r$, since we
only require that $r = e^\eta$ near the outer edge of the grid.
We therefore define
\begin{equation}
r = 2 \sinh\eta + k_1 \eta
\end{equation}
to be the radial coordinate.
Because the throat is described by the first several $\eta$ zones
(see Fig. \ref{fig:clII_misgrid}), 
$k_1$ effectively controls the resolution along and near the throat boundary.

\subsection{The Class III System: \v{C}ade\v{z}--like Coordinates}
\label{sec:classIII}

\v{C}ade\v{z} coordinates (see Fig. \ref{fig:cadezgrid}) are related to
cylindrical coordinates
through the following semi--analytic complex transformation
\begin{eqnarray}
\eta &+& i\xi = \frac{1}{2} \left[\ln \left(z+z_0+i\rho\right) +
\ln \left(z-z_0+i\rho\right)\right] \nonumber \\
&& +\sum_{n=1}^\infty C_n \left[
\frac{1}{(z+z_0+i\rho)^n} + \frac{1}{(z_0-z - i\rho)^n}\right]~,
\label{cadezcoords}
\end{eqnarray}
where $z=\pm z_0$ are the locations of the throat centers.
The constant radial and angular lines lie along the field
and equipotential lines of two equally charged metallic cylinders located
at the centers of the two throats.  Hence the lines of constant $\eta$
are spherical along the throats and in the asymptotic far field,
and a singular saddle point is
introduced midway between the two black holes at the origin $\rho=z=0$.
The coefficients $C_n$ are determined
numerically by a least squares procedure to set the throats (defined
by $\rho_{th}^2 + (z_{th} \pm z_0)^2 = a^2$, where $a=a_1=a_2$ is the
throat radius) to lie on
an $\eta=\eta_0 =$ constant coordinate line.
\vbox{
\figure{
\incfig{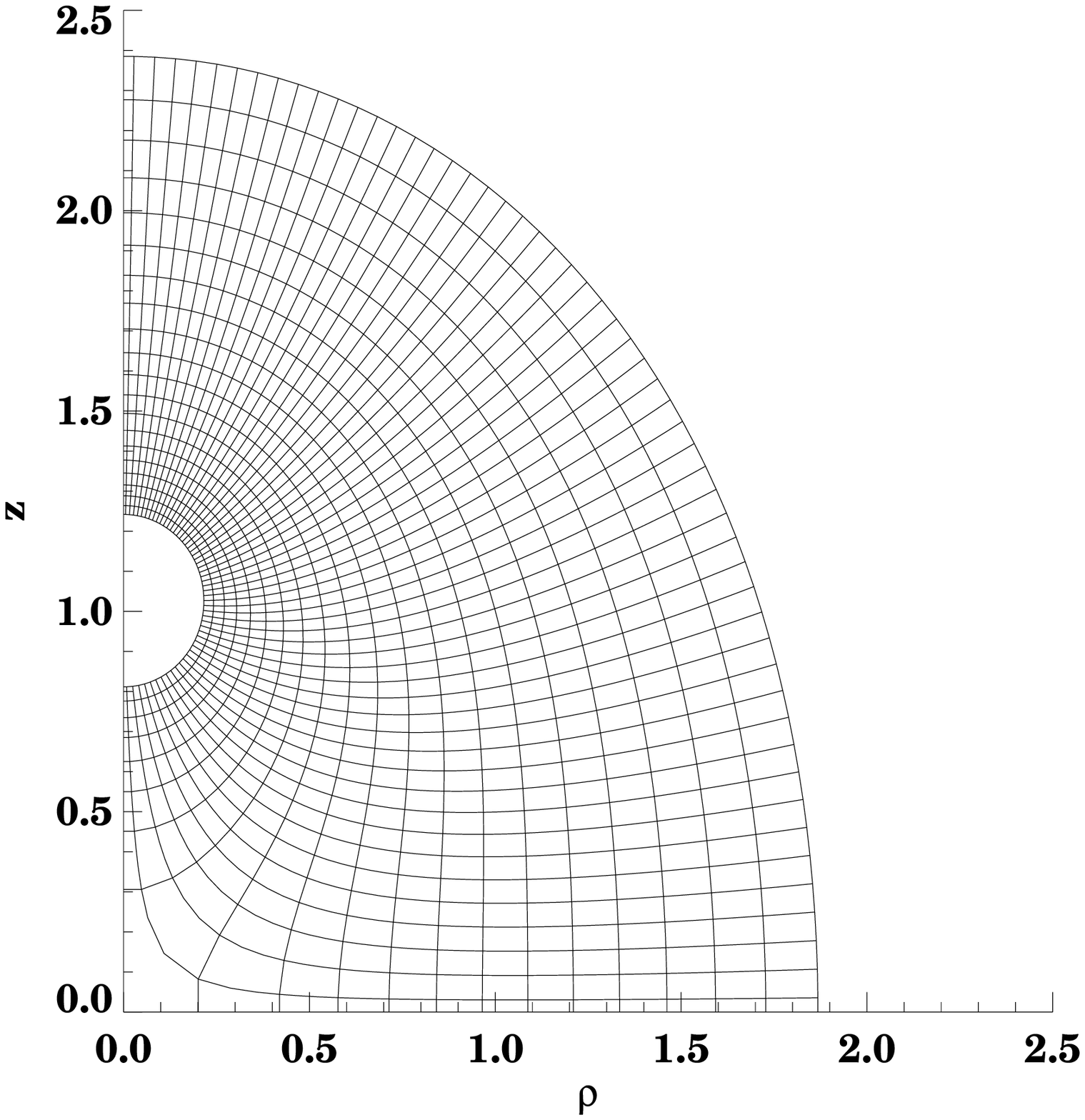}
\Caption{
\v{C}ade\v{z} coordinates for the Misner data with
separation parameter $\mu=2.2$.}
\label{fig:cadezgrid}
}
}

It is relatively easy to generate a \v{C}ade\v{z}--like system using our grid
generation prescription. A radial coordinate is required which
vanishes on the throats (and only on the throats) and asymptotes to
$r$ as $\rho$ and $z$ tend to infinity. Clearly, this can be satisfied with
\begin{equation}
\zeta = \frac{d_1 d_2}{d_1 + d_2} 
        \left(1 - k_1 \exp\left(-k_2\rho^2 - k_3 z^2\right)\right) ,
\end{equation}
where
\begin{eqnarray}
d_1 &=& \sqrt{\rho^2 + (z-z_1)^2}  - a_1 , \\
d_2 &=& \sqrt{\rho^2 + (z+z_2)^2}  - a_2 ,
\end{eqnarray}
and the parameters $k_1$, $k_2$ and $k_3$ are introduced
to control the behavior and resolution of the 
coordinates near the saddle point at the origin.
In contrast with the \v{C}ade\v{z} coordinates, 
we note that the Cauchy-Riemann
relations are not satisfied for this coordinate system.

\section{Generating the Second Coordinate and Other Numerical Issues}
\label{sec:angular}

We have developed two different methods for computing the
second (unspecified) coordinate and to
establish the positions of the $(\eta,\xi)$ grid
points in the cylindrical coordinate $(\rho,z)$ space. 
In addition to evaluating the grid node positions, 
it is also necessary
to compute the Jacobian matrix elements (ie., the coordinate
derivatives) accurately.
The Jacobian matrix and its inverse are needed to transform 
metric functions between the ($\rho$, $z$) and ($\eta$, $\xi$)
spaces and to perform chain rule differentiations of the
cylindrical metric functions on the curvilinear grids
in the coordinate ``patch'' regions described in
Sec.~\ref{sec:applications}.
We turn first to the computation of the Jacobian, and then to
the generation of the second coordinate.

\subsection{Computing the Jacobian}
\label{sec:jacobian}

In order to effectively use the 
curvilinear grids in numerical evolutions,
it is important to generate the Jacobian matrix and its inverse
as accurately as possible, since the gravitational waves
emitted from the collisions are expected to
be very weak in relation to the background geometry. 
Knowing the matrix components amounts to knowing,
at each point on the grids, the coordinate derivatives
$\partial\rho/\partial\eta$, $\partial\rho/\partial\xi$,
$\partial z/\partial\eta$ and $\partial z/\partial\xi$.
A naive approach would be to simply evaluate these derivatives
numerically at each point, after the grid has been calculated,
using a second or higher order discretization stencil.
Instead, we adopt a more accurate
procedure and take advantage of the fact that the derivatives
of the specified coordinate are known analytically. In this
way, only two of the elements need to be computed numerically.

The Jacobian matrix and its inverse can be written as
\begin{equation}
\left [
\begin{array}{cc}
\partial_\eta \rho   & \partial_\eta z \\
\partial_\xi \rho & \partial_\xi z 
\end{array}
\right ]
=
\frac{1}{J}
\left [
\begin{array}{cc}
\partial_z \xi & -\partial_\rho \xi \\
-\partial_z \eta & \partial_\rho \eta
\end{array}
\right ] ,
\label{eqn:jacobian}
\end{equation}
where $J = \partial_\rho \eta \, \partial_z
\xi - \partial_z\eta \, \partial_\rho \xi$
is the Jacobian determinant. One can show from the
orthogonality of the $(\rho,z)$ and $(\eta,\xi)$ coordinates that
the first derivatives obey the Cauchy--Riemann--like conditions
\begin{mathletters}
\label{eqn:cauchy}
\begin{eqnarray}
\partial_z\eta    &=&  \sigma\partial_\rho\xi , \\
\partial_\rho\eta &=& -\sigma\partial_z\xi , 
\end{eqnarray}
\end{mathletters}where
$\sigma = \sqrt{g_{\xi\xi} / g_{\eta\eta}}$,
and $g_{ij}$ is the metric in curvilinear coordinates.
In the class I case, two components of the inverse Jacobian,
$\partial_\rho\eta$ and $\partial_z\eta$,
are known analytically. Thus, using the relationships
(\ref{eqn:jacobian}) and (\ref{eqn:cauchy}), the following class I
elements are determined exactly from the analytic expressions
\begin{mathletters}
\label{eqn:partial_eta}
\begin{eqnarray}
\partial_\eta z &=& \frac{\partial_z \eta}{(\partial_\rho \eta)^2 +
  (\partial_z \eta)^2} , \\
\partial_\eta \rho &=& \frac{\partial_\rho \eta}{(\partial_\rho \eta)^2 +
  (\partial_z \eta)^2} .
\end{eqnarray}
\end{mathletters}The
remaining two elements, $\partial_\xi z$ and $\partial_\xi \rho$,
are approximated by finite differencing the grid data
using a centered fourth order stencil.
In the class II case, the corresponding derivatives of the
coordinate $\xi$ are known, and the expressions analogous to
Eqs. (\ref{eqn:partial_eta}) are derived simply by replacing
$\eta$ with $\xi$.
Additionally, in order to carry out the coordinate patch
evolutions described in Sec.~\ref{subsec:black},
it is also necessary to know the second derivatives of the
coordinates. These are computed in the same manner as the Jacobian:
the specified coordinate derivatives are evaluated analytically,
and derivatives of the second
coordinate are approximated numerically using fourth order stencils.

We note that the orthogonality condition of the angular
and radial coordinates 
\begin{equation}
\frac{\partial z}{\partial \eta}
\frac{\partial z}{\partial \xi}+
\frac{\partial \rho}{\partial \eta}
\frac{\partial \rho}{\partial \xi} = 0 ,
\label{eqn:ortho}
\end{equation}
can be imposed 
here to eliminate a third unknown derivative component 
from the Jacobian matrix, and
reduce the number of numerical discretization operations to one.
However, rather than taking this approach, we defer the
use of the orthogonality relationship to test the accuracy
in computing the coordinate systems and the Jacobian, and
we have confirmed that Eqn. (\ref{eqn:ortho})
converges to zero with the order of the integration method.

\subsection{Line Integration in the $(\rho,z)$ Plane}
\label{sec:line}

One method to generate the second coordinate is to treat 
the lines of constant values as field lines of the known or
specified coordinate. These field lines are computed by integrating
along the normals to the iso--lines, e.g. along the gradient of 
the specified coordinate function.
In other words, given a functional form for one of the coordinates,
the second orthogonal coordinate, $\xi$ in the class I case,
is determined by integrating the following equations
\begin{mathletters}
\label{eq:ortholine}
\begin{eqnarray}
\left.\frac{d \rho}{d \lambda}\right|_{\xi}
       &=& \frac {\partial_\rho \eta}{|\nabla \eta|} , \\
\left.\frac{d z}{d \lambda}\right|_{\xi}
       &=& \frac {\partial_z \eta}{|\nabla \eta|} ,
\end{eqnarray}
\end{mathletters}along 
lines of constant $\xi$.
Here $\lambda$ is an arbitrary integration parameter, and the
normalization factor $|\nabla \eta|$ is introduced to keep the
step sizes regular at points where $\eta$ has large gradients.
This procedure also follows for the class II coordinates,
except that the integrated (differentiated)
coordinate is $\eta$ ($\xi$).
These lines (the second coordinate) are thus guaranteed 
to be orthogonal to the specified coordinate lines.

Because the first coordinate is known exactly,
we can form the spatial gradients analytically, and the problem
reduces conveniently to a straight--forward ODE integration.
In practice, Eqs.~(\ref{eq:ortholine}) are solved using a 4th order
Runge--Kutta integration with a small fixed step size. 
The only difficulty with this method comes with finding the $(\rho,z)$
values at a particular $(\eta,\xi)$ position, since the ODE integrations
may overshoot the destination grid nodes. However,
the step size is chosen small enough (typically about 1\% of
the width of a single grid zone) that we can
interpolate linearly between the staggered points
without sacrificing accuracy.
Once these integrations are completed, the coordinate derivatives are
evaluated using the techniques described in Sec.~\ref{sec:jacobian}.

\subsection{Line Integration in the $(\eta,\xi)$ Plane}
\label{sec:pde}

An alternate, and far more efficient, approach to constructing the
numerical grids involves integrating the 
grid equations (\ref{eqn:partial_eta}),
or their class II equivalents
\begin{mathletters}
\label{eqn:partial_theta}
\begin{eqnarray}
\partial_\xi z &=& \frac{\partial_z \xi}{(\partial_\rho \xi)^2 +
  (\partial_z \xi)^2} , \\
\partial_\xi \rho &=& \frac{\partial_\rho \xi}{(\partial_\rho \xi)^2 +
  (\partial_z \xi)^2} ,
\end{eqnarray}
\end{mathletters}
in the ($\eta$, $\xi$) plane. 
That is, rather than supplying a starting $(\rho,z)$
value, evaluating $\eta$ (or $\xi$) and its gradients, 
and then tracing $\xi$ (or $\eta$) along orthogonal lines
in the ($\rho$, $z$) plane,
we pre--suppose a regular ($\eta$, $\xi$) grid. For class I coordinates,
we supply the $(\rho,z)$ values along the outer boundary
$\eta=\eta_{max}$ which are consistent with a constant
$r$ surface in spherical coordinates, i.e.,
\begin{mathletters}
\begin{eqnarray}
\rho &=& {\rm exp}(\eta_{max}) \sin\theta , \\
z &=& {\rm exp}(\eta_{max}) \cos\theta ,
\end{eqnarray}
\end{mathletters}
and $\xi = \theta$ which is evenly discretized
along $\eta = \eta_{max}$. Eqs.(\ref{eqn:partial_eta})
are then integrated inwards towards the throats.
For the class II coordinates with equal mass black holes, 
we supply the $(\rho,z)$ values along the equator
and integrate Eqs. (\ref{eqn:partial_theta}) towards the axis
and throats. In this case, we use
\begin{mathletters}
\begin{eqnarray}
\rho &=& 2 \sinh\eta+k_1\eta , \\
z &=& 0 ,
\end{eqnarray}
\end{mathletters}
where $\eta$ is evenly discretized from $\eta=0$ to $\eta_{max}$.
Because the derivatives of the specified coordinates are known 
analytically, this integration scheme reduces to a set of
ODEs that we solve using a second order Runge--Kutta scheme with a
step size smaller or equal in size to the
($\eta$, $\xi$) grid spacing used in the evolutions.
As a check on the accuracy of solutions,
the resulting cylindrical coordinate values can be substituted
into the analytic expression for $\eta(\rho,z)$ (class I) or $\xi(\rho,z)$ (class II)
to evaluate the accuracy of the numerical integrations.  We find that they
converge to the truncation order of the integration method.

\section{Applications}
\label{sec:applications}

In this final section we apply the class I grids to numerically
solve the Einstein equations for
the axisymmetric collision of two equal mass black holes,
using the conformal Misner solution for initial data.
Although we have shown how to construct three different grid
classes, we focus here only on the class I type: the class II
evolutions have proven to be less stable than class I, and
class III is similar to the \v{C}ade\v{z} case.
To demonstrate the applicability and accuracy of the new class I grids
to actual evolutions, we repeat the different parameter evolutions
in Papers I--III and compare our results with the published calculations.

\subsection{Initial Data and Evolutions}
\label{subsec:black}

The Misner data set is an axisymmetric and
time--symmetric, single parameter family of solutions 
with the conformally flat spatial 3--metric
\begin{equation}
dl^2 = \Psi^4 \left[d\rho^2 + dz^2 + \rho^2 d\phi^2\right],
\label{eqn:metric_cyl}
\end{equation}
where $\rho$ and $z$ are the cylindrical coordinates, and
\begin{equation}
\Psi = 1 +\sum_{n=1}^\infty \frac{1}{\sinh (n\mu)} \left[
       \frac{1}{\sqrt{\rho^2 + \left(z+\coth n\mu \right)^2}} + 
       \frac{1}{\sqrt{\rho^2 + \left(z-\coth n\mu \right)^2}}\right]~.
\end{equation}
The conformal factor $\Psi$
solves the Hamiltonian constraint with the proper
isometry imposed between the upper and lower sheets and represents two
equal mass, non-rotating black holes aligned along the axis of symmetry
(the $z$--axis), and centered at $z=\pm\coth\mu$ with radius
$a=1/\sinh\mu$.  The free parameter $\mu$ defines the 
total or ADM mass of the spacetime
and the proper distance along the spacelike
geodesic connecting the two throats.  Increasing $\mu$ decreases the
total mass of the system and sets the two holes further away from one
another.

For general axisymmetric transformed coordinates, the
3--metric (\ref{eqn:metric_cyl}) can be written as
\begin{equation}
dl^2 = {\Psi_C^4}  \left[
       {\cal J} \left(\rho_\eta^2 + z_\eta^2\right)~d\eta^2
     + {\cal J} \left(\rho_\xi^2  + z_\xi^2 \right)~d\xi^2  
      +\left(\frac{\rho^2}{\sin^2\xi}\right) {\cal J}~\sin^2\xi~d\phi^2
      \right],
\label{eqn:metric_cur}
\end{equation}
where $\Psi_C = \Psi/{\cal J}^{1/4}$, ${\cal J}$ is
a regularization variable that can be used to simplify
the metric components and to provide a stabilizing element
in the numerical evolutions,
($\eta$, $\xi$) are the logarithmic radial--like
and angular--like curvilinear coordinates, and the $\sin^2\xi$
term is explicitly factored out of the $g_{\phi\phi}$ component
partly for historical conventions
and partly to help regularize the numerical evolutions:
We evolve the conformal metric and extrinsic curvature
components as described in Paper II. 
However, there ${\cal J}$ was defined
to be the Jacobian determinant of the coordinate transformation
(${\cal J} = J$) so that $g_{\eta\eta} = g_{\xi\xi} = 1$ initially. 
Here we simply set ${\cal J} = 1$ since there is no obvious
advantage to renormalize one of the components when
the Cauchy--Riemann conditions are not satisfied.
The partial coordinate
derivatives in (\ref{eqn:metric_cur}) are computed as described
in Sec.~\ref{sec:jacobian}, completing the initial data.
The data is then evolved according to the same procedures
described in Paper II, ie.,
using the maximal slicing condition
for the lapse function with antisymmetric boundary conditions
at the throat surfaces,
and an elliptic condition for the shift vector
to preserve the metric in curvilinear coordinates to be
diagonal throughout the evolution.

To provide more stable evolutions, especially for the
widely separated black hole cases, we utilize a
coordinate ``patch'' as described in Paper II.
Evolutions in this patched domain, which covers the
saddle points and portions of the axis, are performed using the
cylindrical coordinate based metric components
and chain rule derivatives to compute spatial gradients
across the curvilinear grid nodes. The solutions are then
transformed using the general tensor relations to
reconstruct the curvilinear metric and
extrinsic curvature components, and then linearly blending the
results into the rest of the spacetime, which is evolved
in a normal manner using the curvilinear metric components
on the curvilinear grid.
The coordinate patch used in Papers I--III, although
only several zones deep in the angular direction, extends from the
throat all the way out to the outer boundary in the radial direction.
For comparison (see Fig. \ref{fig:patches}), the domain in the
class I coordinates which requires a patch is localized to just the
first few zones in the radial direction. 
As a result, when the black holes
merge to form a common event horizon, the patched coordinates
and the associated numerical evolutions
become irrelevant as the lapse collapses to zero over this region
before the metric shear grows enough to
disrupt the solutions. The evolutions with class I coordinates
are therefore more robust and less sensitive to patch parameters
then previous calculations.
\vbox{
\figure{
\incfig{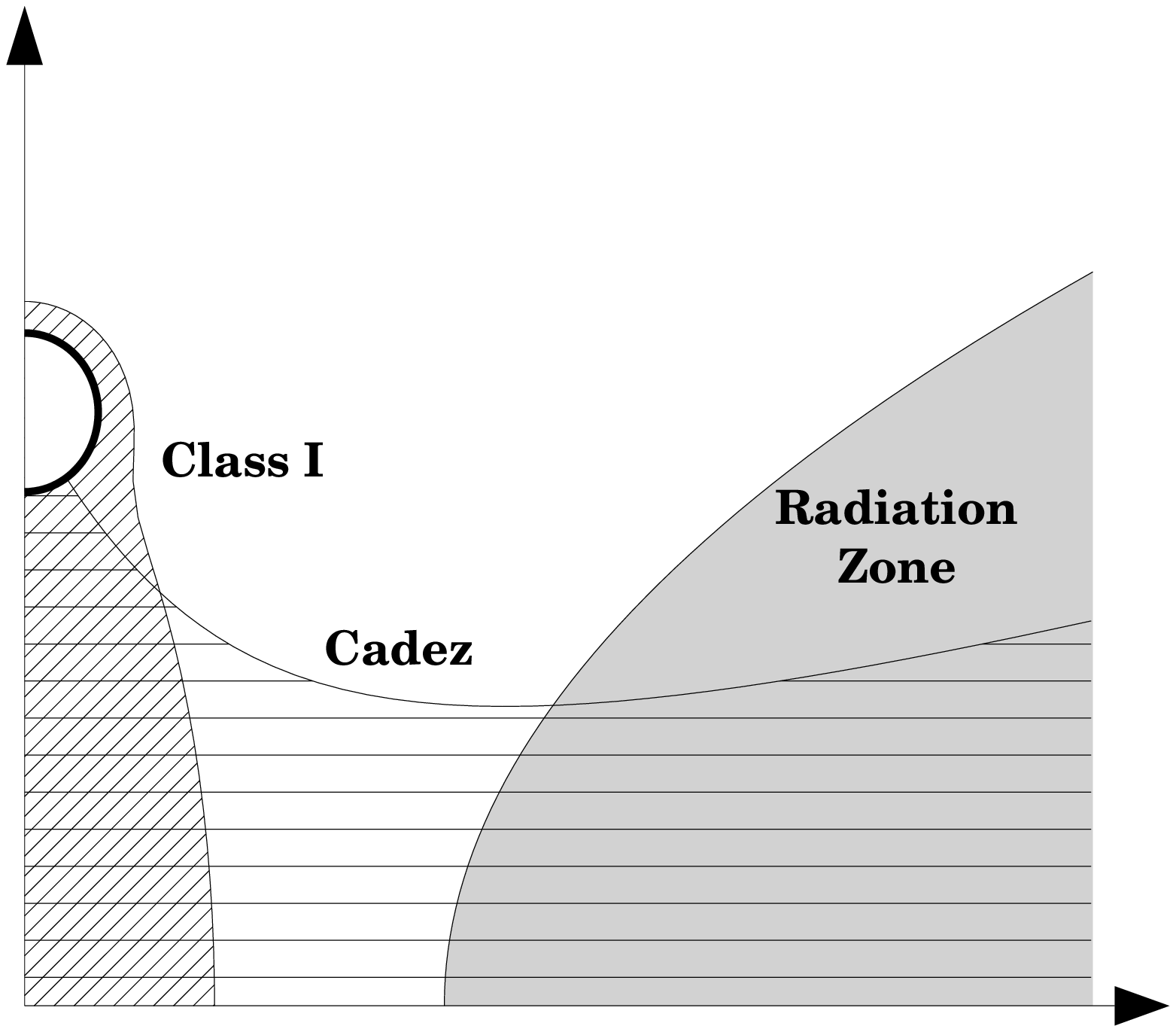}
\Caption{Locations, shapes and sizes of the coordinate patches used in
the Misner $\mu = 2.7$ grids for the \v{C}ade\v{z} 
(horizontally hatched domain) and class I 
(diagonally hatched domain) cases.
The shaded region labeled ``radiation zone'' represents the domain into
which most of the gravitational radiation is emitted. That this zone 
is concentrated along the equator is attributed to the axisymmetric 
nature of the collision. Despite
the representative shape of the radiation zone, the actual
radiation extraction is performed on 2--spheres starting at a radius of
30$M$ (= 15$M_{ADM}$) from the origin.
}
\label{fig:patches}
}
}

\subsection{Results}
\label{sec:results}
\vbox{
\figure{
\incfig{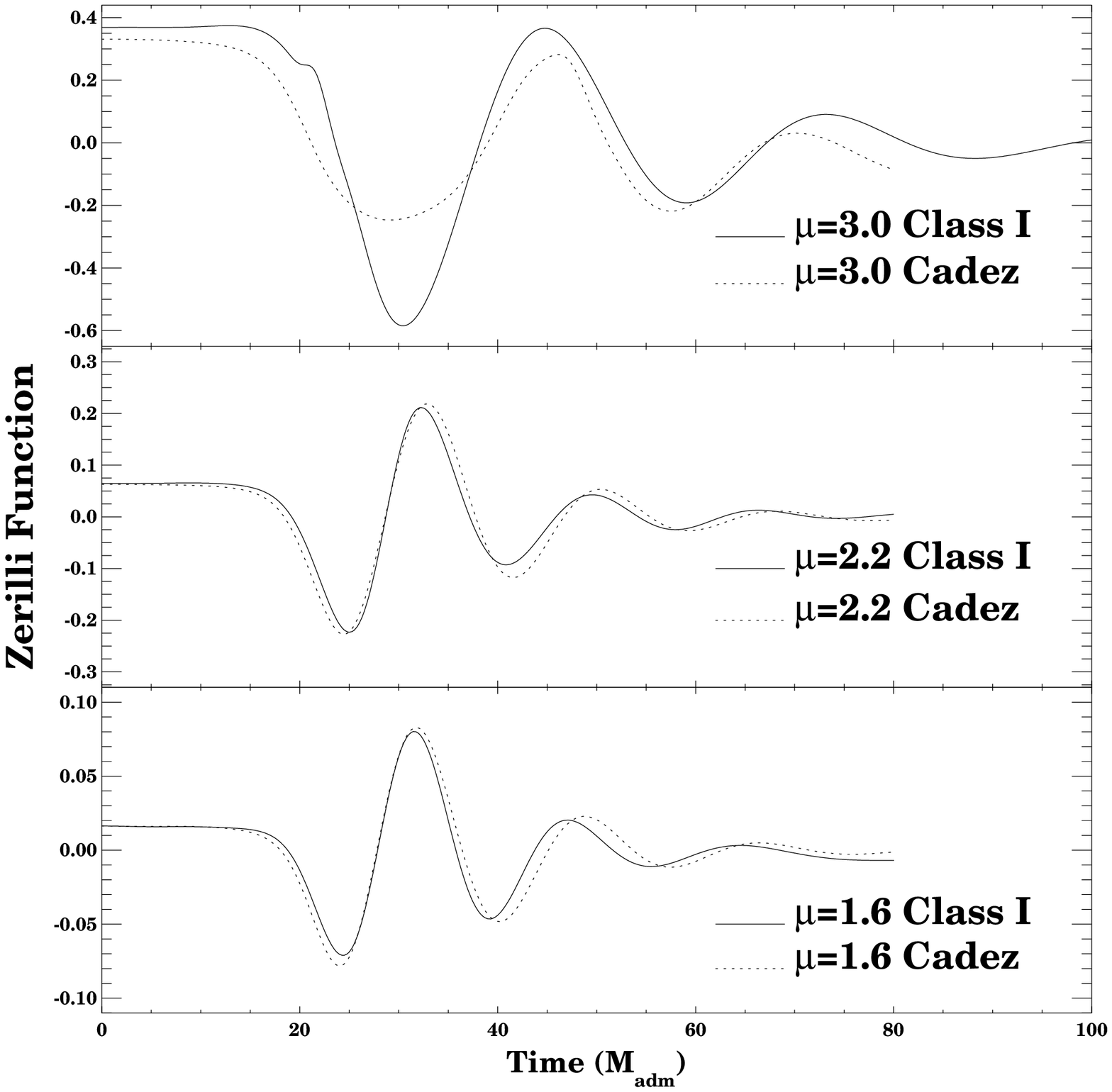}
\Caption{
Gravitational waveforms emitted from the head--on 
collision of two black holes evolved with the \v{C}ade\v{z} and class I
grids for the $\mu =$ 1.6, 2.2 and 3.0 Misner data.
The waveforms are
extracted at a radius of 30$M$ from the origin, where
$M = M_{ADM}/2$ is the approximate single black hole mass.
The evolutions in each case are performed at a grid resolution
of 200$\times$35 (radial$\times$angular) cells.
}
\label{fig:gwave1}
}
}

We first consider three different cases with Misner parameters
$\mu =$ 1.6, 2.2, and 3.0, and compare the class I
gravitational waveforms with previous results using the \v{C}ade\v{z} grid.
Each calculation is performed at the same grid resolution
using 200$\times$35 radial$\times$angular zones.
The Misner parameters
are chosen so that the smaller value represents a data set
in which the two black holes are already merged within a single
event horizon, and the subsequent evolution is that
of a single distorted black hole ringing
down to the Schwarzschild solution as it emits 
gravitational waves. The data from the
larger Misner parameter cases correspond to two
distinct black holes separated by proper distances
of $L = 8.92 M$ ($\mu = 2.2$) and 
$L = 15.8 M$ ($\mu = 3.0$) between the two throats,
where $M$ is half the ADM mass of the spacetime
(or approximately the mass of a single black hole).
Comparison plots of the dominant $\ell=2$ 
Zerilli waveforms, extracted at a 
distance of 30$M$ from the origin, are shown in
Fig. \ref{fig:gwave1} for the three different Misner cases.
We find {\it maximum} relative differences less than
about 10\% in amplitude and 4\% in phase for the 
$\mu = $ 1.6 and 2.2 cases (with absolute deviations
$< 10^{-2}$ in amplitude), 
and up to about 200\%
and 4\% differences in amplitude and phase for the
more difficult $\mu = $ 3.0 case in which the black
holes are initially highly separated.

To emphasize the (in)stability of the solutions at late times
when the signal crossing the detector becomes weaker,
we plot in Fig. \ref{fig:gwave_log} the logarithm of the
absolute value of the Zerilli function for the relatively
uncertain $\mu=3.0$ case. Notice the
class I grid solution maintains a more regular
oscillatory behavior and consistent damping rate throughout
the wave signal and for longer periods of time
than the \v{C}ade\v{z} case, which begins to break
down at about 70$M_{ADM}$.

\vbox{
\figure{
\incfig{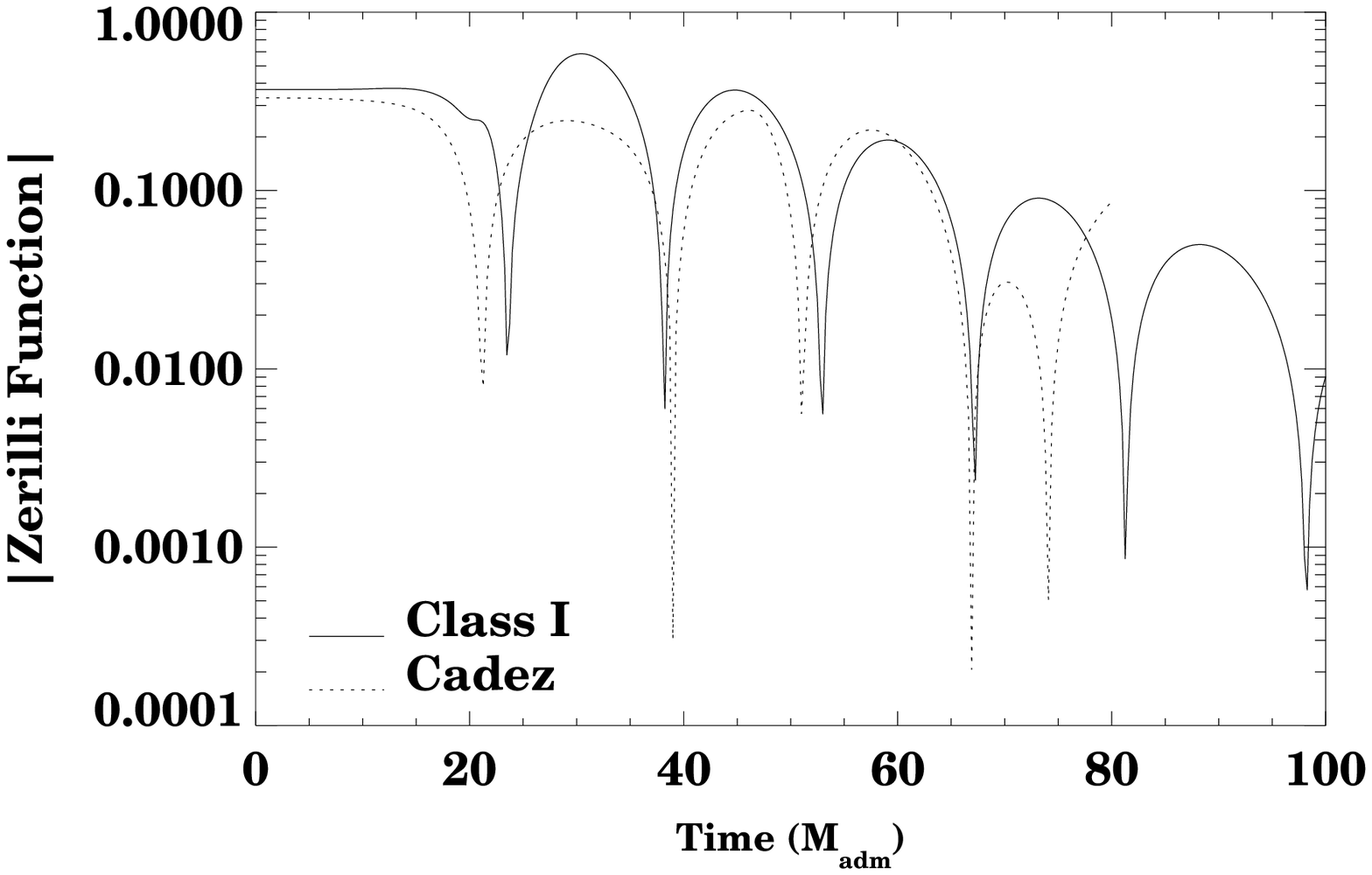}
\Caption{
The absolute value of the Zerilli function is shown
for the $\mu = 3.0$ case using the 
class I and \v{C}ade\v{z} grids.
The logarithmic scale highlights the oscillations in the waveform
and the exponential damping of the wave amplitude.
Although the waveforms compare favorably at 
smaller values of $\mu$, there is a significant 
difference and improvement
in the regularity of oscillations in the class I system
compared with the less stable \v{C}ade\v{z} case for this
large initial separation data.
}
\label{fig:gwave_log}
}
}

Next, we reproduce in Fig. \ref{fig:radiation} the equivalent
of Fig. 14 in Paper III. The total $\ell=2$ radiated energy 
(in units of the ADM mass, $M_{ADM} = 2 M$) emitted during the black
hole collisions is plotted as a function of the 
initial separation distance (in units of $M$) between the two throats.
Also included in the figure are results from Paper III,
the Davies et al. \cite{DRPP} (DRPP) point particle 
calculation ($E = 0.0104 m^2/M$, plotted for
$m=M=M_{ADM}/2$), and its reduced mass correction
($m \rightarrow mM/(m+M)$).
Results from the class I and \v{C}ade\v{z} grid evolutions match
extremely well, better than 3\%, for separation parameters $\mu \le 2.2$
corresponding to initial physical separations of $L \le 8.92 M$,
and improve significantly at smaller $\mu$ values.
However, deviations between the class I and \v{C}ade\v{z}
results become greater for black holes with further
initial separations, reflecting the difficulty in evolving
highly separated black holes for long periods of time.

To assert a measure of uncertainty in our results, we perform
several different calculations of the $\mu=2.7$ and 3.0 data,
varying the grid resolution, patch width, and coordinate
parameters as defined in Sec. \ref{sec:classI} to manipulate the shape
of the coordinate lines. For the more 
problematic \v{C}ade\v{z} evolutions,
additional parameters include the patch length, duration
and diffusion, as well as varied treatments of the coordinate
singularity (for example, shift vector specifications,
discretization stencils, and regularization of certain
metric and curvature components). The symbols in
Fig. \ref{fig:radiation} represent the median results and
the error bars indicate the variance with computational
parameters. The variances are similar to the
differences observed between the
\v{C}ade\v{z} and class I results: roughly 30\% and 100\%
in the $\mu = 2.7$ and 3.0 cases respectively, with a trend for better
agreement with increased resolution (though the resolution
studies are limited by the axis instability). For the
$\mu \le 2.2$ cases the variances are too small to plot, but are
consistent with the observed agreement in the evolutions.
Considering the sensitivity
of the results to grid and patch parameters,
and the difficulties in evolving these systems using maximal slicing
conditions and in axisymmetry, the overall results
agree fairly well. Furthermore, the numerical
calculations are in reasonable agreement with the
reduced mass approximation and the semi--analytical calculation
of Ara\'ujo and Oliveira \cite{AO}
in the large separation limit.

\vbox{
\figure{
\incfig{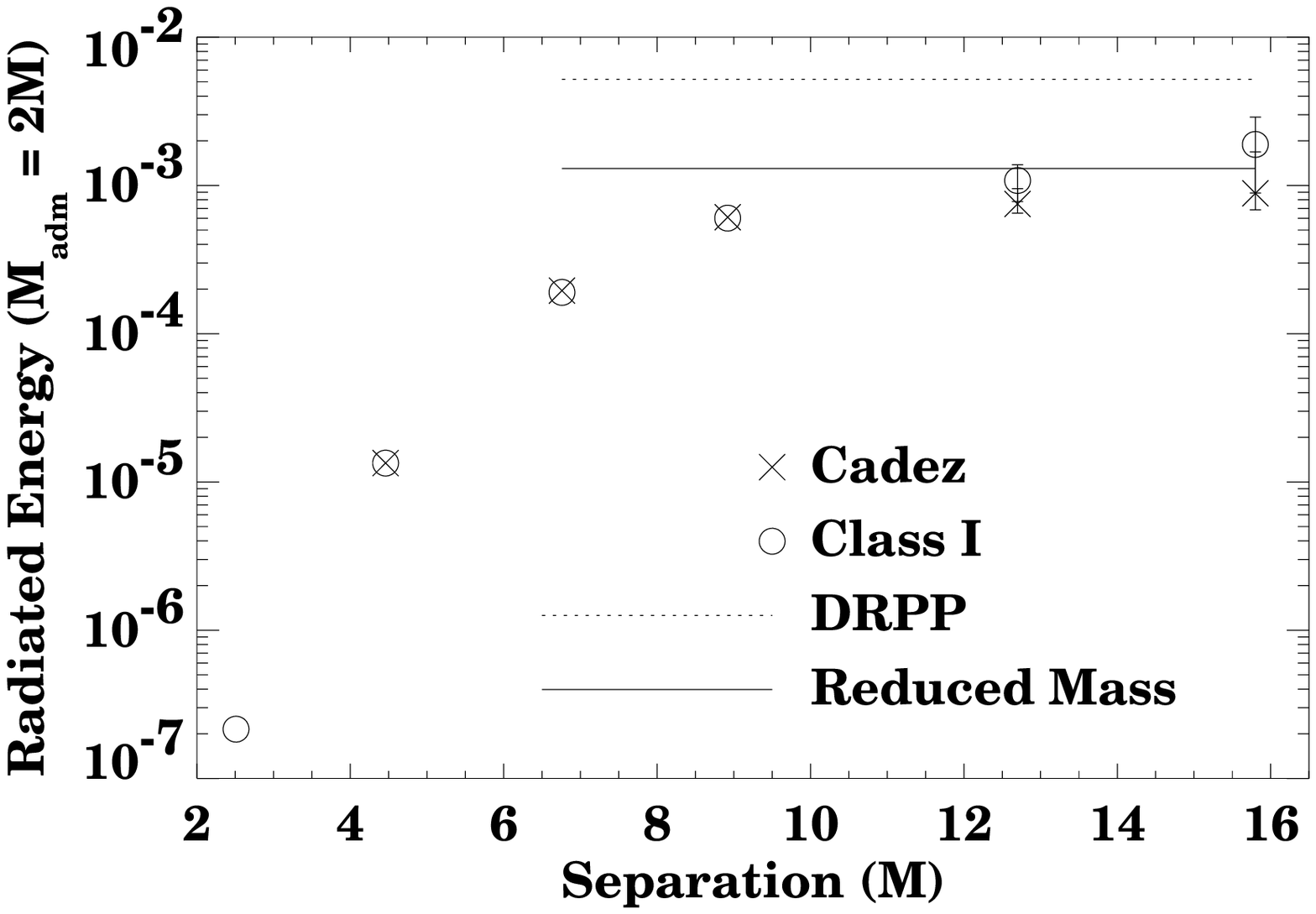}
\Caption{
Total gravitational wave energy 
(in units of the ADM mass)
emitted from the head--on collision of two
equal mass black holes as a function of the 
separation distance between the two throats
(in units of $M=M_{ADM}/2$).
The class I and \v{C}ade\v{z} grid
results are plotted together for comparison, along with
the DRPP \protect\cite{DRPP} calculation of a particle
falling into a more massive black hole, and
its reduced mass correction.
The error bars represent the uncertainties estimated
by performing the evolutions for different
computational parameters as described in the text.
The uncertainties in both the class I and \v{C}ade\v{z}
evolutions are comparable to differences
between the two different grid results, ranging from
$<3$\% for separations $L<10 M$, 30\% for
$L\sim 13 M$, and 100\% for $L\sim 16 M$.
\label{fig:radiation}
}
}
}

Fig. \ref{fig:radiation} also demonstrates an added advantage of
the new grid generation procedure to construct numerical
grids in the very low Misner parameter cases. Due to 
convergence problems in the Newton--Raphson
iterative inversion of Eqn. (\ref{cadezcoords}) when the
black hole throats are placed too close to the saddle point,
previous calculations were limited to $\mu\ge 0.7$ \cite{Price}.
The closest separation data shown in Fig. \ref{fig:radiation}
corresponds to $\mu=0.5$ (or $L=2.51 M$), although grids for
even smaller values of $\mu$ can be easily generated.

An additional benefit from these new coordinates is their regularity
at the origin, which makes calculations of the event horizon and null
generators more accurate as the black holes merge.  In Fig.
\ref{fig:ehembed} we show the evolution of the embedding of the event
horizon found in the $\mu=2.2$ case evolved with the class I
coordinates. The embedding of the horizon is smooth (except at the
cusps on the $z$--axis), as were previous embeddings of the horizon in
this spacetime \cite{Matzner95a}.  However, the null surfaces on the
class I grid contain not only the horizon (in the domain $\eta > 0$),
but also naturally contain the locus of generators waiting to find the
horizon, as discussed in \cite{Libson94}. More precisely, to generate
Fig. 11 of \cite{Libson94} a coordinate transformation from
\v{C}ade\v{z} to $(\rho,z)$ coordinates was required. This coordinate
transformation is not needed to evolve the locus in class I
coordinates.  This is more than a convenience, however. Since the null
surface consisting of the horizon plus locus is naturally represented
as a smooth continuous surface in Class I coordinates, the entire null
surface can be unambiguously embedded in 3--space, which was not
possible with the \v{C}ade\v{z} grid.  This allows us to determine the
separation between the horizons in embedded space using some relevant
physical measure.  While in \v{C}ade\v{z} coordinates the horizon
separation was determined to keep the outer surface of the ``pair of
pants'' figure smooth, here we determine the separation by embedding
the entire null surface. This gives a natural separation between the
horizons from the embedding of the locus and determines the geometry
by the horizon.  We use this separation to place the holes in the
``wristwatch'' (or ``pair of pants from above'')
Fig.~\ref{fig:ehembed}. The class I coordinates also allow for more
detailed examinations of the caustic line at the coalescence point,
which will be discussed elsewhere.

\vbox{
\figure{
\incfif{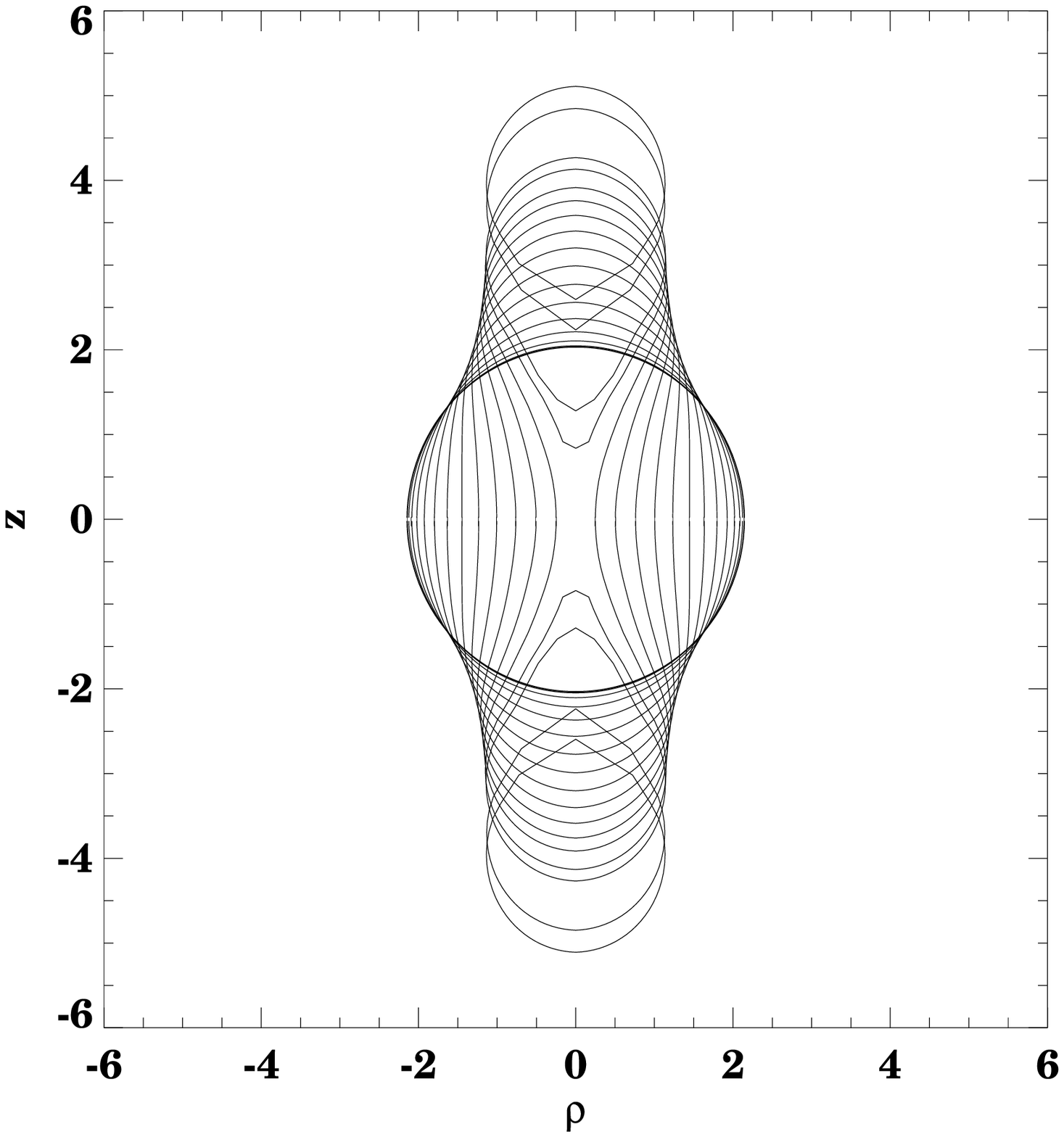}
\Caption{Embedding of the event horizon for the $\mu = 2.2$ 
Misner data with the class I evolution.
The geometry of the class I coordinates allows the separation
between the holes in this figure to be physical, not artificial as in
previous two black hole collision embedding diagrams derived
from Cadez coordinates.
Also, the regularity of the class I grid at the origin allows
for more accurate examinations of the caustic line
at the coalescence point.
}
\label{fig:ehembed}
}
}

\section{Conclusions}
\label{sec:conclusion}

In many ways the axisymmetric evolution of two colliding black holes
has been stalled due to the lack of a coordinate system without
pathological instabilities or grid singularities. 
By developing new techniques
to generate alternative grids, and by creating 
body--fitting grid geometries with singularities
on the throats of the two black holes, we are able to achieve more
stable long time evolutions of two black hole systems. 
We have demonstrated the applicability of these new grids
in actual numerical evolutions of 
Misner's initial data set for the head--on collision of two equal
mass black holes.
The calculations presented here confirm the existing
results to fairly good accuracy in the 
restricted stable parameter range,
with maximum relative deviations less than 3\% 
in the radiated energies for the
$\mu \le 2.2$ cases corresponding to initial proper separation
distances of $L \le 8.92 M$ (with significantly smaller deviations
for the lower $\mu$ cases).
The differences increase to about
30\% for $\mu=2.7$ ($L = 12.7 M$)
and 100\% for $\mu=3.0$ ($L = 15.8 M$), which are comparable
to the uncertainties in the evolutions as defined in
Sec. \ref{sec:results} for both the \v{C}ade\v{z} and class I
systems.
These differences reflect the difficulty in evolving
black holes for long periods of time with maximal
slicing conditions, and the sensitivity of the evolutions
to treatments of the axis instability and coordinate singularities.

Basically
the \v{C}ade\v{z} and class I evolutions yield consistent results,
even at the high $\mu$ values, especially when considering the 
changes observed in waveforms and energies by
varying the computational parameters.
In addition to confirming previous results, 
and because no specialized treatment of the coordinate
singularities is required, it seems promising
that we can evolve physical systems with these new
coordinates which were not as easily
addressable with previous codes using
either the \v{C}ade\v{z} or cylindrical coordinate systems.
For example, we are currently extending this work to
evolve spinning black hole and 
unequal mass black hole collisions.

To conclude, we emphasize the following advantages of these new
coordinates:
(1) 
they achieve better zone coverage in the strong field interaction region
near the origin -- caustics, photon generators and embeddings of
the event horizon are better resolved;
(2) 
the coordinate patch extends over a much 
smaller domain in the class I system -- just a few zones
radially, and localized to the $z$--axis -- 
so it is not an unstabilizing element at late times;
(3) 
the resulting gravitational
waveforms from evolutions are not as sensitive to the patch parameters,
such as its width, length, duration and diffusion parameters;
(4)
due to its robustness and lack of a need for specialized
treatments of the saddle points,
the new code is more simplified and
easily generalizable to include non-equal mass black
holes and spinning black hole collisions;
and (5)
the new coordinates allow
a larger range of initial data in the Misner parameter to be
evolved, including evolutions of black holes that are
further separated (though the accuracy is questionable
for $\mu \ge 3.0$), and more closely spaced
(for smaller order perturbations) than in previous calculations.
In addition,
the new coordinates offer all the same advantages as the
\v{C}ade\v{z} coordinates: They are logarithmic in the radial
direction, and spherical on the throat
and in the asymptotic wave zone, thus allowing for the same
simplified treatment of boundary conditions and waveform extraction.

\acknowledgements

It is a pleasure to thank Bernd Br\"ugmann, Greg Daues, 
Joan Mass\'o, Sam Oliveira, Bernard Schutz, Ed Seidel,
John Shalf, Wai--Mo Suen, Malcolm Tobias,
and especially Scott Klasky for many discussions.
PA would also like to thank members of the Albert--Einstein--Institut
for their hospitality and support during which part of this work
was carried out. The computations were performed on the Origin 2000
at the Albert--Einstein--Institut and the 
National Center for Supercomputing Applications, and 
the C90 at the Pittsburgh Supercomputing Center.



\end{document}